\documentclass[journal]{IEEEtran}

\usepackage{cite}

\ifCLASSINFOpdf
\else
\fi

\usepackage{amsmath}

\usepackage{verbatim}
\usepackage{cite}
\usepackage{xcolor}
\usepackage{bm}
\usepackage{multirow}
\usepackage{booktabs}
\usepackage{caption}
\usepackage{subcaption}
\usepackage{chngcntr}
\usepackage[at]{easylist}
\usepackage{array}
\usepackage[hyphens]{url}
\usepackage{hyperref}
\usepackage[hyphenbreaks]{breakurl}

\ifCLASSINFOpdf
   \usepackage[pdftex]{graphicx}
\else
   \usepackage[dvips]{graphicx}
\fi

\newcolumntype{C}[1]{>{\centering\let\newline\\\arraybackslash\hspace{0pt}}m{#1}}
\begin{document}

\title{California Test System (CATS): A Geographically Accurate Test System based on the California Grid}

\author{Sofia Taylor$^*$,~%
        Aditya Rangarajan$^*$,~%
        Noah Rhodes,~%
        Jonathan Snodgrass,~%
        Bernie Lesieutre,~%
        Line A. Roald,~%
\thanks{Sofia Taylor, Aditya Rangarajan, Noah Rhodes, Bernie Lesieutre and Line A. Roald are with the Department of Electrical and Computer Engineering, University of Wisconsin, Madison, WI, ZIP USA. Jonathan Snodgrass is with the Department of Electrical Engineering at Texas A\&M University. E-mail: smtaylor8@wisc.edu. \emph{$^*$The two first authors Sofia Taylor and Aditya Rangarajan contributed equally to this work.}}%
\thanks{This work is funded in part by the Power Systems Engineering Research Center (PSERC) through project S-91, the National Science Foundation (NSF) under Grant. No. ECCS-2045860, and the NSF Graduate Research Fellowship Program under Grant No. DGE-1747503.}
}

\maketitle

\begin{abstract}
This paper presents the California Test System (CATS), a synthetic transmission grid in California that can be used by the public for power systems policy research without revealing any critical energy information. The proposed synthetic grid combines publicly available geographic data of California's electric infrastructure, such as the actual locations of transmission corridors, with invented topology and transmission line parameters that are ``realistic but not real". The result is a power grid test system that is suitable 
for power flow and policy analyses with 
geo-referenced applications, including studies related to weather, topography, and socio-economic considerations.
The methods used to develop and evaluate the CATS grid are documented in detail in this report.
\end{abstract}

\IEEEpeerreviewmaketitle

\section{Introduction}
\begin{figure}[t]
    \centering
    \begin{subfigure}[t]{0.36\columnwidth}
        \centering
        \includegraphics[width=\columnwidth]{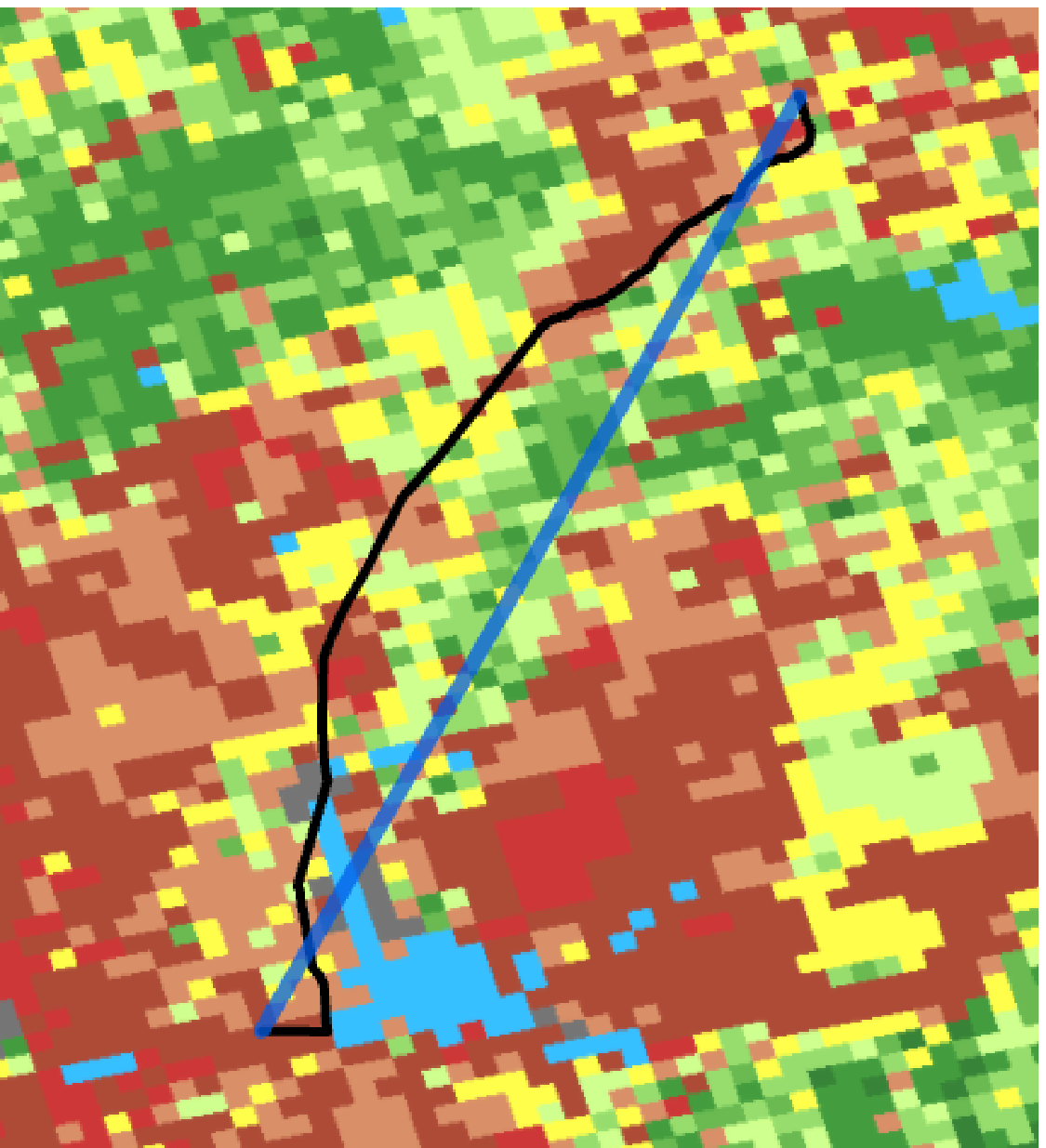}
        \caption{\small}
        \label{fig:fire_line}
    \end{subfigure}
    \begin{subfigure}[t]{0.615\columnwidth}
        \centering
        \includegraphics[width=\columnwidth]{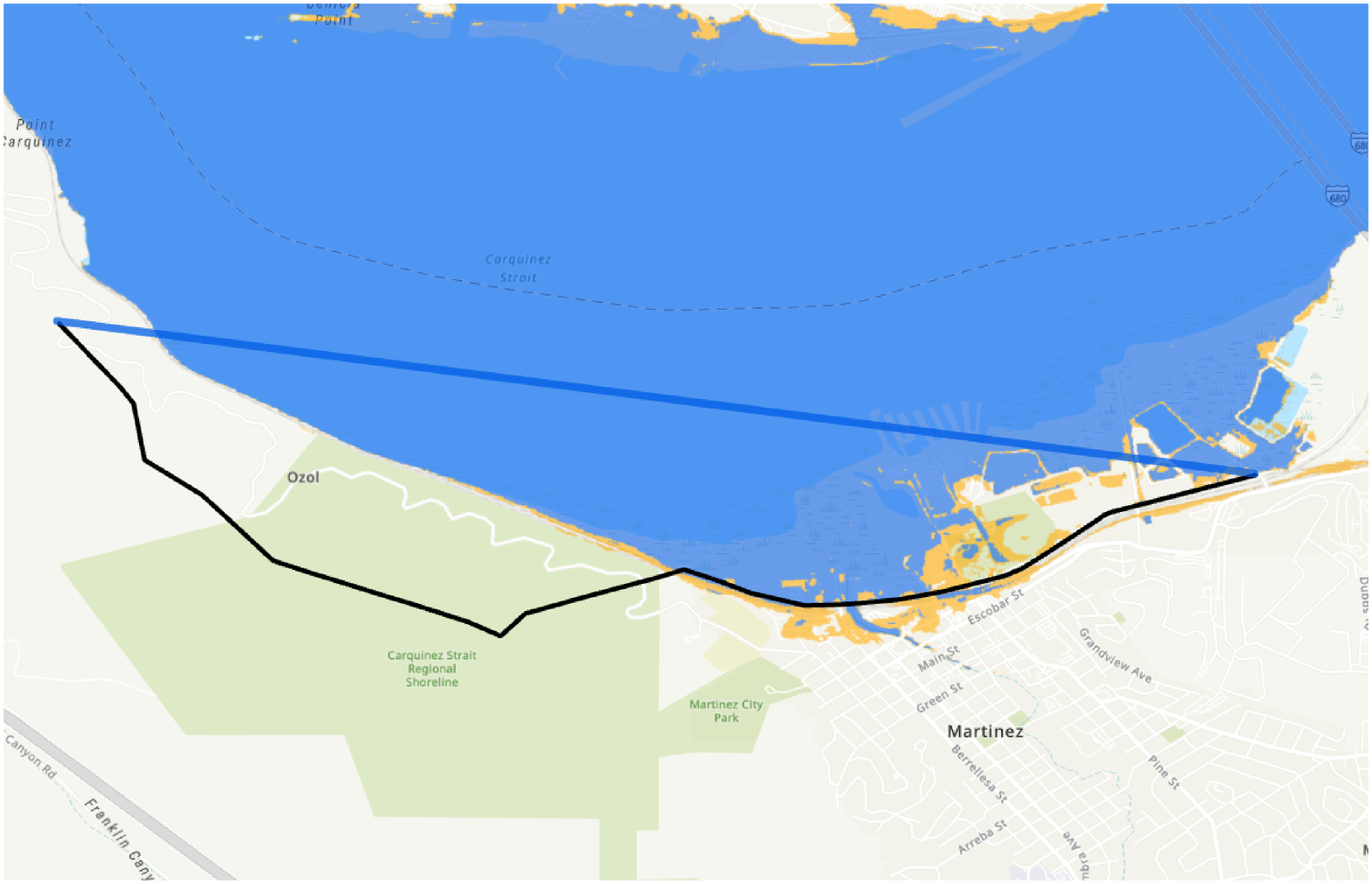}
        \caption{\small}
        \label{fig:slr_line}
    \end{subfigure}
    \caption{\small Images depicting the usefulness of the accurate paths of power lines (in black) versus their point-to-point approximations (in blue) in analyses with geo-referenced data. Fig. \ref{fig:fire_line}  shows a 230kV line in northern California overlaid on a wildfire risk map. Fig. \ref{fig:slr_line} shows a 60kV line in the Bay Area overlaid on NOAA 5-foot sea level rise projections.}
    \vspace{-3mm}
    \label{fig:geo_accuracy}
\end{figure}

Effective polices are essential for operating a reliable, resilient, equitable, and economical electric grid. Examples of existing regulations and polices include grid reliability standards \cite{NERC_Standards}, procedures for operating fair electricity markets \cite{FERC_Tariff}, and state and regional policies for considering and approving new facilities and rates. For resiliency,
there is growing recognition that we must prepare the grid for more frequent extreme weather events, including increased wildfire risk, more destructive hurricanes, tornadoes, and flooding, and heat waves that may elevate electric load.
Modeling and understanding these impacts and assessing relevant policies requires us to consider not only the electric grid, but also the environmental and social context around it. Developing new procedures for operations and planning that capture this context requires access to geographically accurate grid data that can be correlated with other public data sources, such as data on wildfire and flooding risks, and  information on the vulnerability of the population to power outages and weather impacts. 
Fig. \ref{fig:geo_accuracy} demonstrates the importance of geographical accuracy by comparing two transmission lines in California with their linear, point-to-point approximations. The transmission line in Fig. \ref{fig:fire_line} is overlaid on top of wildfire risk data, showing that the actual line path passes through areas of much higher wildfire risk than the straight line path.
In Fig. \ref{fig:slr_line}, we can see that the straight line approximation is not useful for a study related to sea level rise, as it passes through a body of water rather than curving around the coastline.

 The most natural model for analyzing geographically-, environmentally-, and socially-dependent policies is the actual grid model. Nevertheless, there are important reasons to keep the real electric grid model from public view. 
For example, details about electricity generation, transmission, and distribution might compromise the competitiveness and privacy of market operations.
Additionally, such knowledge could facilitate cyber and physical attacks on electric infrastructure, which pose a threat to public health and national security. 
To fortify the privacy and security of electric infrastructure in the United States, the Federal Energy Regulatory Commission (FERC) has deemed certain power system information to be Critical Energy/Electric Infrastructure Information (CEII) \cite{FERC_2021}. As a result, CEII data and models of power systems are not publicly available. Researchers that obtain permission can access CEII data, but can not publish it along with test case results, creating challenges for benchmarking and sharing research.

Thus, to fullfill research needs, researchers have developed synthetic grid models with realistic but artificial parameters. Synthetic grids include a set of electrical components, including transmission lines, transformers, buses, substations, generators, and loads, with defined parameters and connections that determine how current will flow through the system. The classic IEEE test cases were among the first synthetic grid models \cite{IEEE_Test_Cases}, followed by recent synthetic networks created by Texas A\&M University \cite{TAMUelectricgrids} and University of Wisconsin-Madison \cite{UWelectricgrids}, the Reliability Test System -- Grid Modernization Lab Consortium (RTS-GMLC) network \cite{Barrows2019}, and other test cases for benchmarking AC optimal power flow (OPF) algorithms \cite{pglib-opf-2019}. These synthetic grids are typically easy to use relative to real grid models and allow users to freely share results.

One shortcoming of current synthetic grids is their lack of realistic geographic information. Though inspired by real electric grids, %
the IEEE test cases do not include any geographic information. In contrast, the recently developed grids \cite{TAMUelectricgrids,UWelectricgrids,Barrows2019} all contain geographic coordinates that span various regions across the United States. These models do not represent any particular real power system, but model the graph characteristics and electrical behavior of real power systems \cite{BirchfieldXu2017,BirchfieldSchweitzer2017,Sogol2021,Barrows2019}. Thus, the transmission lines, substations and buses do not correspond to any existing power equipment. 
Also, the transmission lines are represented as straight lines connecting two nodes, rather than nonlinear paths that curve based on local topography, vegetation, and property ownership. 

 As a concrete example of the shortcomings of current grid data, recent works \cite{rhodes2021balancing}\cite{taylor2022framework} consider policies to mitigate the risk of wildfires induced by the power grid, but are limited by the lack of a grid model with realistic geography. 
The model in \cite{rhodes2021balancing} balanced the competing risks associated with wildfire ignition from power equipment and preemptive power shutoffs.
The model, which includes constraints to model power flow (PF), is tested on the RTS-GMLC \cite{RTSGMLC}, a synthetic grid with straight power line paths that largely traverse the desert and thus have relatively low wildfire risk. Grid data with more realistic transmission line paths would be useful in this case.
The framework in \cite{taylor2022framework} optimally selects power lines to harden to mitigate wildfire ignition risk. This model is tested on geographic data (without PF parameters) of the actual power line paths in California \cite{CECdata}. However, because a complete grid model (with component connections and parameters) of this data did not previously exist, it was not possible to enforce PF constraints in this test case.

Thus, there is a need for synthetic grids without CEII-protected information that accurately represent the geography of real power systems and are suitable for PF analyses. This paper aims to address this need.

The main contribution of this paper is the California Test System (CATS), a geographically accurate synthetic transmission network model that is located in the state of California. The starting point for this model is geographic data of California's electric infrastructure, which is publicly available through the California Energy Commission (CEC) \cite{CECdata} and the Energy Information Agency (EIA) \cite{EIA860}. However, while there is an abundance of publicly available data, there are several important parts of the data that are missing. This includes the topology of the system (e.g., connections between components), transmission line and generation parameters, and locations and parameters of transformers and reactive power compensation devices. %
Thus, we supplement the available geographical grid information with the necessary synthetic data to create a geo-located test system suitable for policy studies that utilize PF and OPF analysis. To achieve this, we add approximate substation and node topologies to the system, 
leverage additional generation and load data from the EIA \cite{EIA860} and California Independent System Operator (CAISO) \cite{CAISO_prod_2019},
and assign realistic line parameters based on publicly available data from FERC \cite{FERC_Form1}. The result is an open-source, non-CEII transmission network model suitable for geo-located policy study applications and large scale PF and OPF studies. The CATS grid model is available in a GitHub repository in MATPOWER and GIS formats \cite{githubRepo}.

In summary, the contributions of the paper are twofold. First, we provide a new, openly available test system for PF and OPF studies that allow for large-scale analysis and correlation with other geo-referenced data. The test system has been evaluated for a full year of load and generation data, and is made easily accessible to researchers and the general public through our GitHub repository \cite{githubRepo}. 
Second, we describe the procedure that we used to create the CATS grid model. While the description primarily serves to explain how this specific grid model was developed, the procedure can be adapted to create similar transmission and distribution models in other geographic locations.

The rest of this paper is organized as follows. 
Section \ref{sec:data} details the data sources and Sections \ref{sec:Topology}-\ref{sec:Reactive-Power} describe the methodology used to create the test system.
Section \ref{sec:Metrics-Evaluation} presents evaluation metrics and performance results. 
Section \ref{sec:geo-conclusion} summarizes the contributions, limitations, and future work.

\section{Data Sources}
\label{sec:data}
We use several sources of publicly accessible data to develop the CATS synthetic grid 
including electric infrastructure geographic information, generation data, load profiles, and transmission line parameter data. This section describes the sources for these data. We describe methods that use this data to create CATS in Sections \ref{sec:Topology}--\ref{sec:Reactive-Power}.

\subsection{California Electric Infrastructure Geographic Data}
Geographic information systems (GIS) use maps to represent spatial data. 
We obtain GIS data of California's power lines and substations, including locations and attributes, from the CEC \cite{CECdata}.
While the CEC provides GIS data of California's power plants, the EIA also publishes information about California's power plants, including geographic coordinates, in the Form EIA-860 \cite{EIA860}. The location and capacity of the generators and plants from the CEC and EIA sources are similar, but not exactly the same. To inform generator locations in the CATS network, we select the 2019 Form EIA-860 data because it provides additional fields for the generators.

\subsection{Generation Data}
\label{sec:Gen-Data}
In addition to geographic coordinates, the 2019 Form EIA-860 contains useful generator attribute information, such as the fuel and unit types, power factor, and minimum and nameplate capacity MW values. However, it does not contain necessary details about renewable energy generation output and non-renewable generation cost curves, so we use 2019 state-wide renewable data published by CAISO \cite{CAISO_prod_2019}. For nuclear generators, the cost is estimated from 2019 expenditures \cite{EIA_expenditures} and production \cite{CEC_generation}. Quadratic cost curve coefficients for all other generation are obtained from \cite{Sogol2018}, and the cost of electricity imports is estimated from 2019 EIA data \cite{EIA_expenditures}.

\subsection{Load Data}
To generate load profiles, we leverage publicly available aggregate hourly load data from CAISO for 2019 \cite{CAISO_prod_2019} as an input to a method developed as part of the EPIGRIDS project \cite{EPIGRIDS}, which disaggregated state-wide temporal load profiles to individual bus-level loads. This method produced hourly load data at census tract level granularity that captures geographic variation throughout California \cite{Wang2020,snodgrass2021tractable}. %

\subsection{Transmission Line and Transformer Parameter Data}
\label{sec:TL-Parameter-Data}
Transmission line parameters, including resistance (R), reactance (X) and susceptance (B), are protected data and not available in public data sets. These parameters are necessary for grid analysis, so we use publicly available data to  generate realistic (but synthetic) parameters for the transmission lines and transformers in the CATS network. FERC publishes historical and current annual reports of information about electric utilities in the United States. Two of these reports, the Form No. 1 ``Annual Report for Major Electric Utility'' \cite{FERC_Form1}, and the Form No. 715 ``Annual Transmission Planning and Evaluation Report'' \cite{FERC_Form715}, contain useful data for assigning line parameters. The Form 1 is publicly available, while the Form 715 is CEII and no longer released to the public. Although access to the Form 715 can be requested, we opt instead to only use the Form 1 from 2010 and previously published average values and statistical data derived from the Form 715 \cite{snodgrass2021tractable, BirchfieldSchweitzer2017} to avoid concerns regarding protected data in the CATS grid model. 

The Form 1 was created for utilities to report their bulk electric system assets to FERC for annual accounting. In California, we use data from three investor-owned utility companies (Pacific Gas \& Electric, Southern California Edison, and San Diego Gas \& Electric) from the FERC Form 1 report  to create the proposed grid. 
The data include voltage level, transmission line length, number of conductors per phase, conductor size and material, transmission structure material, and construction type.

The Form No. 1 does not contain any useful data for transformer parameters. We therefore leverage average per unit impedance values using the transformer base MVA for each pair of primary-secondary voltages from \cite{snodgrass2021tractable}. We obtain the X/R ratios for all transformers, with MVA values ranging from 50 MVA to 2000 MVA, from \cite{BirchfieldSchweitzer2017}.

\section{Grid Topology}
\label{sec:Topology}
While the data from CEC and EIA provide accurate information about the geographic location of individual components, they do not provide a description of how the components are connected, i.e., they lack important information regarding the system topology. To create a fully connected network suitable for power system simulation and analysis, we  introduce a method for connecting components and describe various data cleaning and processing steps.

\subsection{Initial Topology}
The available datasets include substation locations, generator locations, and transmission line paths. The first step in achieving a connected topology is to assign substation connections to the end points of the transmission lines. To do this, we calculate the distance from a transmission line endpoint to each substation in the network, assign the closest substation as a connection, and repeat for each transmission line endpoint in the network. Similarly, for assigning generators to substations, we calculate the distance from each generator to each substation and assign a generator connection to the closest substation. 
While this connectivity method is simple and intuitive, several data challenges cause the resulting network to be a poor representation of the CAISO network.
We discuss these challenges and their solutions below.

\subsection{Connecting Transmission Lines to Substations}
The transmission lines from the CEC data represent the actual line locations, but the data does not accurately model electrical circuits and connections. For example,
many transmission lines are represented as several line segments. These line segments should not be connected to substations, but rather to each other. If we force all line ends to connect to a substation, many of the segments are connected incorrectly. %

To remedy this challenge, we create
additional nodes to represent electrical interconnections between transmission line segments. We place these ``added nodes'' at every transmission line segment endpoint that does not already have a node or substation within a small search radius. A conservative search radius of just 12 meters ensures that nodes are still placed at the endpoints of very short line segments that exist within cities. This procedure adds thousands of nodes to the grid, but creates a much more accurate topology.

\subsection{Transmission Lines that Branch into Separate Paths}
\begin{figure}[t]
    \centering
    \includegraphics[width=0.4\columnwidth]{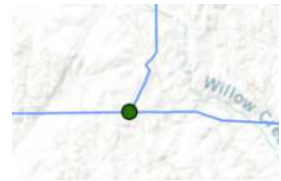}
    \caption{\small Transmission line branching off a from another line.}
    \label{fig:tap_line}
\end{figure}

In the GIS data, it is common for transmission lines to have other lines branching away along their length. An example is shown in Fig. \ref{fig:tap_line}. Per the method described above, we add a new node at the branching line's endpoint. However, the main line often does not terminate at the node, and therefore is not modeled as an electrical connection. In this case, we must segment the main line into two parts to correctly model the connection to the branch.
Thus, we segment all such lines into smaller line paths to create the appropriate end points and connections between the circuits. Unfortunately,  the process creates a few hundred extremely short line segments, some smaller than one meter. We manually delete or merge these lines segments with another line.

\subsection{Manually Resolving Topology Issues} \label{resolve_topology}
\begin{figure}[t]
    \centering
    \includegraphics[width=0.85\columnwidth]{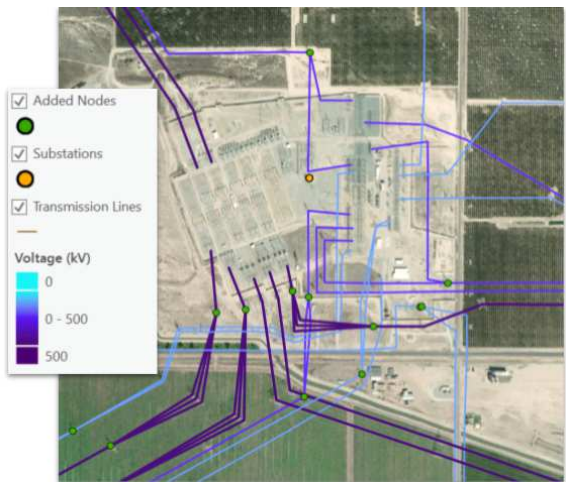}
    \caption{\small Transmission lines at the Midway Substation in Kilowatt, California, overlaid on top of a Google Maps satellite image.} 
    \label{fig:mid_sub}
\end{figure}

After the above steps are complete, we run the network connectivity process (where all lines are connected to the closest node) again.
However, although the above methods resolve a lot of issues, they also cause secondary topology problems that require manual analysis and data cleaning.
Certain parts of the topology, such as Midway Substation in Kilowatt, California, shown in Fig. \ref{fig:mid_sub}, are not correctly connected because too many nodes have been added to nearby transmission line endpoints (due to the conservative search radius used when adding new nodes). Large substations like Midway cover a large area of land in reality but are represented as a single point, with relatively large distances between the substation point and adjacent line endpoints. As a result, our methods add many extra nodes, and many lines connect to nodes outside of the substation instead of to the substation itself. %
To ensure that the lines connect correctly to the substation point, we manually remove extra nodes in the area around the Midway substation.

To aid in identifying other locations where there may be connectivity challenges, we create a tool to calculate the graph-distance (with line length as edge weights) and geographic-distance between node pairs. A large discrepancy between these values could indicate that nearby nodes are not electrically connected when they should be. This allows for easier identification of locations similar to Midway substation, where creating correct topology connections in a fully automated way is challenging. After using the tool to identify a possible connectivity issue, we manually modify the geographic data as needed.

One particular connectivity challenge arises when our methods assign the same node connection to both ends of a line. In these cases, an added node at an interconnection is often closer than the substation node, and therefore both line endpoints are assigned to the same node. To address this, we produce a list of these lines and manually fix them by deleting nodes and merging, extending, and deleting lines in ways that allow components to correctly connect.

\subsection{Substation Transformers}
The final topology challenge is the addition of transformers. The datasets include line voltage levels, but not connections between voltage levels. We must create substation topologies if multiple voltage levels connect to the same substation. Thus, when multiple line voltage levels connect at the same node, we split the node into multiple nodes with transformers in between them and then reassign line and generator connections.

We create new nodes connected by transformers as follows. 
\begin{enumerate}
    \item Identify the number of voltage levels $V$ at a node. 
\item Add $V-1$ additional nodes to the network. 
\item Add new transformer branches to connect the nodes.
\item Assign each line that connected to the original node to the new node representing the correct voltage level.
\item Assign generators to the highest voltage level in the set.
\end{enumerate}

\vspace{5pt}

For each of the topology creation steps, we modify and clean some of the input data and then re-run the automated connectivity steps. 
In Section \ref{sec:gen_load_data}, we discuss our process for assigning generation and load data to this topology. 
Then, Section \ref{sec:connect} concludes the topology creation phase by describing the final grid connectivity steps.

\begin{figure*}
    \centering
    \vspace{-2.5em}
    \begin{minipage}[b][][b]{.5\textwidth}
        \centering
        \hspace{-2em}
        \includegraphics[width=1.05\linewidth]{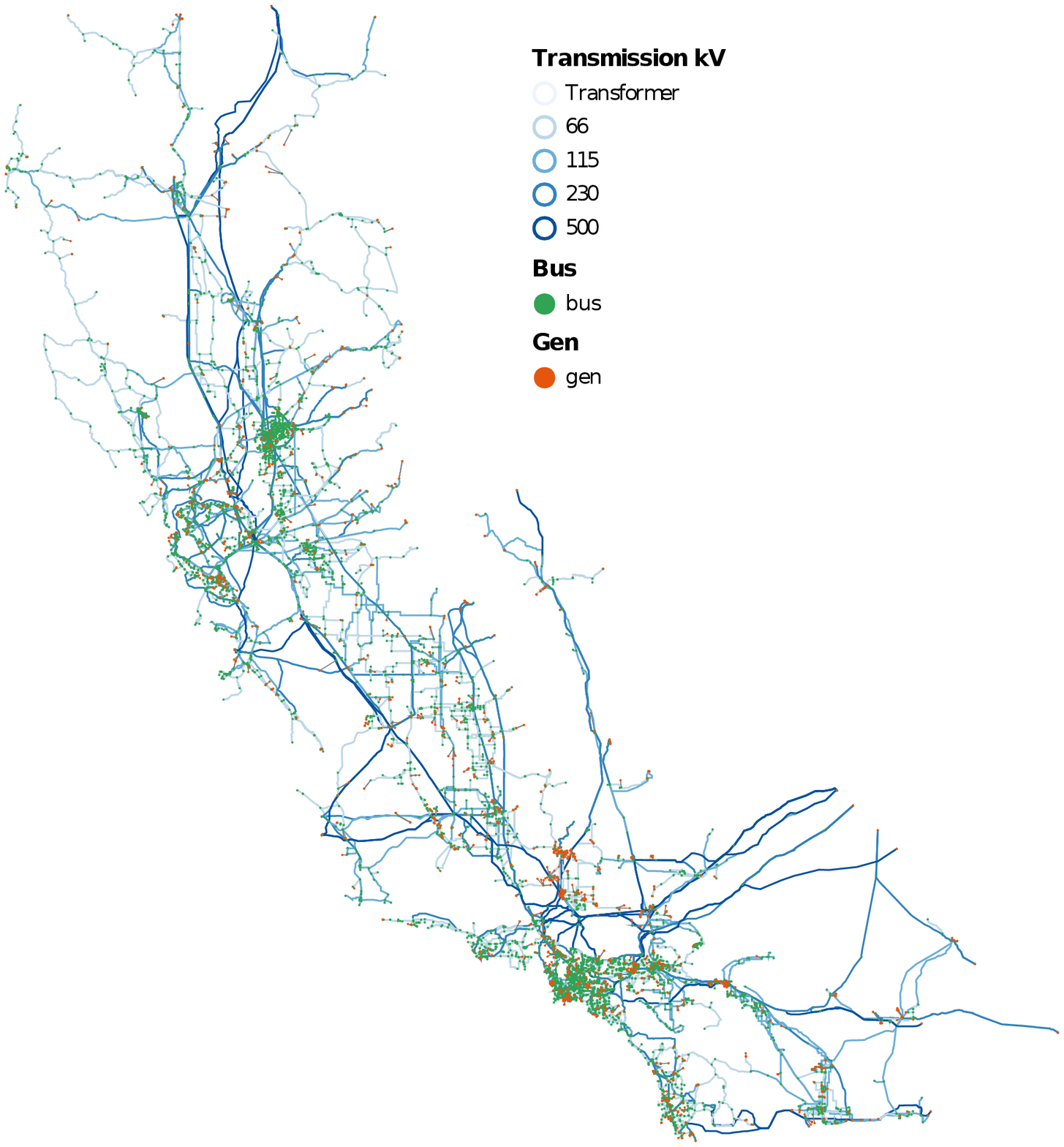}
        \caption{\small California Test System topology.}
        \label{fig:calgrid_topology}
    \end{minipage}%
    \begin{minipage}[b][][b]{0.5\textwidth}
        \centering
        \includegraphics[width=0.7\linewidth]{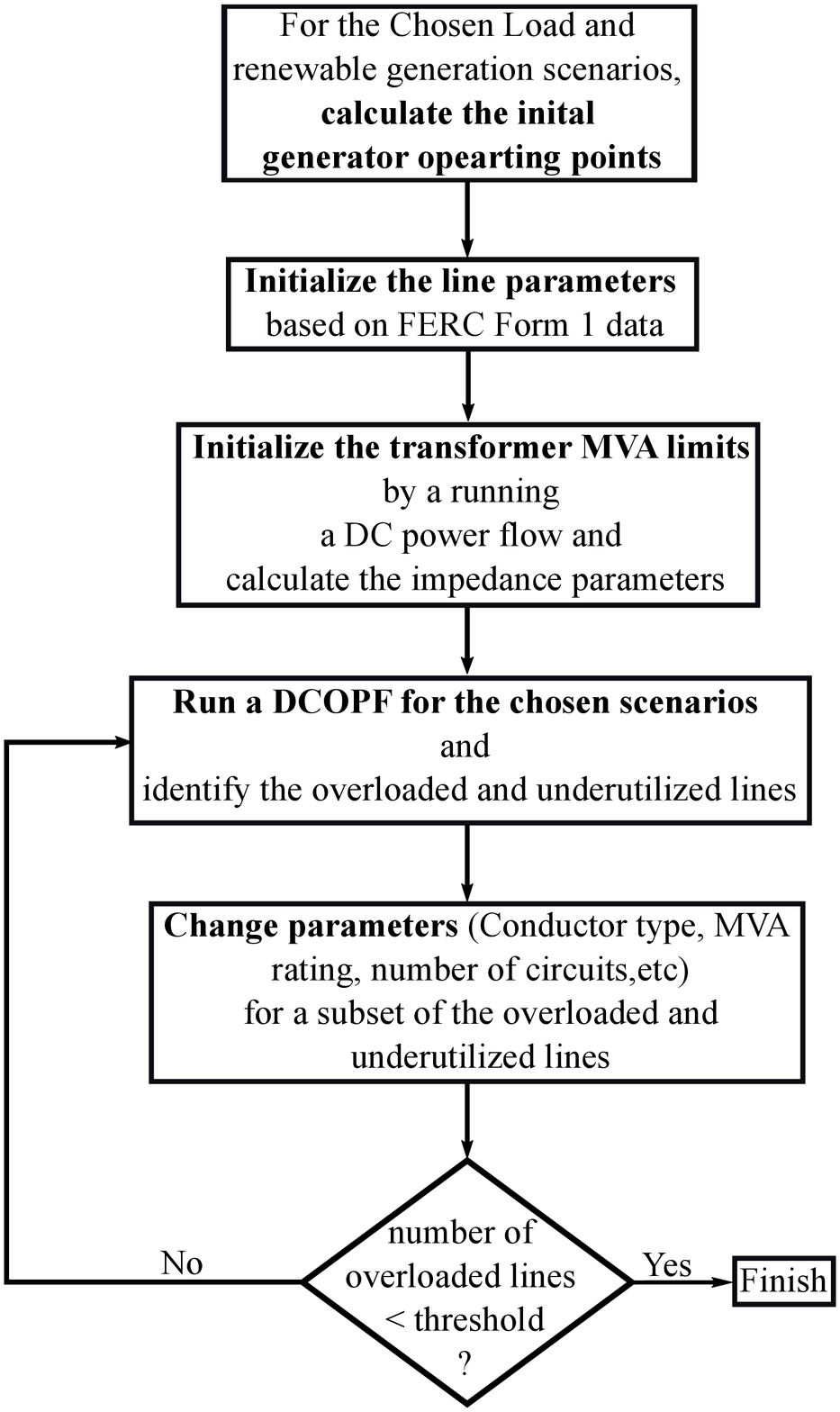}
        \caption{\small Flowchart of the process for assigning transmission line and transformer parameters.}
        \label{fig:flow}
    \end{minipage}
\end{figure*}

\section{Assigning Generator and Load Data} \label{sec:gen_load_data}
Next, we describe our methodology for assigning generation, load and renewable energy data in the CATS. 

\emph{Assigning Data for Conventional Generators:}
The Form EIA-860 files contain static attributes, including the nameplate capacity, nameplate power factor, latitude, and longitude.
In addition to the generator locations and capacities, we also need cost curves to model generator dispatch within the electricity market. 
We correlate the cost curve coefficients from \cite{Sogol2018}, which is based on the 2010 Form EIA-860 generators, with the 2019 Form EIA-860 generators in our system. For `Plant Codes' and `Generator ID's' that directly match between the 2010 cost curve data from \cite{Sogol2018} and the 2019 generators, the cost curve coefficients are simply copied over. 
For nuclear generators, we assign a linear cost, 
 which we estimate by dividing the total nuclear generation expenditures \cite{EIA_expenditures} by production \cite{CEC_generation} for California in 2019. We also assign a linear cost 
to import generators, based on EIA expenditure data \cite{EIA_expenditures}.
For the remaining non-renewable generators, we assign the coefficients of the closest-sized generator of the same type.
Generators of types ``Other Natural Gas'' and ``Municipal Solid Waste'' are not present in the data from \cite{Sogol2018}, so we instead approximate cost information based on generators of types ``Natural Gas Steam Turbine'' and ``Landfill Gas'', respectively.
For the renewable generators, we assign coefficients of zero.

\emph{Assigning Data for Renewable Generation:}
To account for the variability of solar and wind generation, we use the state-wide generation data published by CAISO for 2019 \cite{CAISO_prod_2019}. The data consists of 5-minute load and generation from different resources, which we sample at hourly intervals. First, we scale the capacities of all the renewable generators in the CATS grid such that their total capacity matches the total capacity of the CAISO dataset. 
Next, for every scenario considered, we scale the capacity of each solar and wind generator according to its actual rating, as shown in equations \eqref{eq:solarscaling} and \eqref{eq:windscaling}.
\begin{subequations}
\begin{align}
\label{eq:solarscaling}
P_{PV,i}^{max,s} = \frac{P_{PV,i}^{max}}{\sum_{j \in \mathcal{PV}} P_{PV,j}^{max}} P_{PV}^{s}
\quad \forall i \in \mathcal{PV}, s \in \mathcal{S} 
\\
\label{eq:windscaling}
P_{W,i}^{max,s} = \frac{P_{W,i}^{max}}{\sum_{j \in \mathcal{W}} P_{W,j}^{max}} P_{W}^{s}
\quad \forall i \in \mathcal{W}, s \in \mathcal{S} 
\end{align}
\end{subequations}
In the above equations, $\mathcal{PV}$ is the set of all solar generators and $\mathcal{W}$ is the set of all wind generators in the CATS. The set $\mathcal{S}$ denotes the generation scenarios considered. %
$P_{g,i}^{max}$ is the actual capacity of generator $i$, $P_{g}^{s}$ represents the total amount of solar or wind generation in the system, respectively, and $P_{g,i}^{max,s}$ is scaled capacity proportional to the generation from renewable sources in scenario $s$.
We note that the above assignment policy does not contain any geographical variation of renewable generation availability. In future work, we could improve the renewable energy production scenarios to account for more geographic variability.

\emph{Assigning Load:}
We utilize hourly load data scenarios from the EPIGRIDS project \cite{snodgrass2021tractable, Wang2020}. The loads are given per census tract, and therefore need to be assigned to a specific bus in the CATS grid. A first approximation is to assign each load to the nearest CEC substation. However, this method would leave many substations with very high load, many with no load, and some radial substations with no load or generation. For more realistic load assignment, we use an optimization assignment problem, shown in Problem \ref{prob:load_assign}, that minimizes the distance between the EPIGRIDS load locations and the CEC substations. 
\begin{subequations}
    \label{prob:load_assign}
    \begin{align}
        \min_x \quad & \sum_{i \in \cal{L}}\sum_{j \in \cal{N}} x_{ij} c_{ij}\\
        \mbox{s.t.} & \sum_{j \in \cal{N}} x_{ij} = 1 && \forall i \in \cal{L} \label{eq:single_assign}\\
        & \sum_{i \in \cal{L}} x_{ij}\ge 1 && \forall j \in \cal{N} \label{eq:assign_sufficient}%
    \end{align}
\end{subequations}
Here, the load assignment variable $x_{ij}$ is a binary variable that determines if load $i$ is assigned to node $j$. The objective function minimizes the cumulative distance from the location of each load to the substations, represented by $c_{ij}$.
The set $\cal{L}$ is the set of all loads from EPIGRIDS, with one load per census tract. The set $\cal{N}$ is the set of CATS nodes where we require the algorithm to assign at least one load. This set $\cal{N}$ includes the original substations from the CEC dataset, as well as the added nodes located at the end of radial branches that do not have generators attached. 
The constraints on the load assignment are that each load $i \in \cal{L}$ is assigned is assigned to exactly one node \eqref{eq:single_assign}, and that each node $j \in \cal{N}$ is assigned at least one load \eqref{eq:assign_sufficient}. The set of loads $\cal{L}$ is larger than the set of nodes $\cal{N}$ at which we assigned the loads, permitting a feasible solution to the problem. Since this problem satisfies the conditions for total unimodularity, it produces a solution with $x_{ij}\in\{0,1\}$ without explicitly enforcing that $x_{ij}$ is binary.

\section{Grid Connectivity}
\label{sec:connect}

After the above processing steps,
there is no guarantee that the network is a single connected graph. As an example, a power line and two substations located on the Santa Catalina Island do not connect to the mainland power grid. Similarly, substations and transmission lines in the northwest corner of the state do not directly connect to the rest of the California power system, and instead connect through the state of Oregon.

To ensure a single power grid network, we select the largest network from the many sub-networks. Over 92.8\% of the buses, generators, loads and transmission lines are contained in the single largest network, providing a good representation for the majority of the power grid within the state. The final grid topology is shown in Fig. \ref{fig:calgrid_topology}.

\section{Assigning Line and Transformer Parameters}

To obtain a system which gives rise to feasible and realistic PF and OPF solutions, we must create realistic line and transformer parameters, such as impedance and MVA limits. This section describes our procedure for generating these limits. Key steps are outlined in Fig. \ref{fig:flow}. %

\subsection{Input Data}

\subsubsection{Load and Generation Scenarios}

To create a realistic power system network model that is feasible for a wide range of load and renewable generation scenarios, we must consider more than one loading condition when assigning line parameters. 
Specifically, we choose a total of 245 scenarios. We choose the first 121 scenarios to be the hour with maximum load and the 60 hours before and after. We choose the next 121 scenarios to be the hour with the minimum load and the 60 hours before and after. In addition, we consider 3 scenarios that represent the hour with the maximum solar generation, the hour with maximum wind generation and the hour with the lowest overall renewable generation. 
This ensures that we consider a range of both load and renewable generation scenarios, as well as different hours of the day and days of the week.

\subsubsection{Initial Generation Profiles for Each Scenario}
When creating generation profiles for the scenarios, we want to reflect typical operating conditions (i.e., conditions that allow the lowest cost generators to run). At the same time, we should avoid over-optimizing the grid such that transmissions lines are sized only to support the lowest cost generator dispatch (economic dispatch). Also, electricity demand changes throughout a day or week, as well as across several months or years. If the grid creation process does not exhibit enough variety in the generator dispatch and unit commitment, the grid will not be able to support varying load flow patterns.

Past experience has also shown that synthetic grid models are highly sensitive to the set of generators committed, and that each generator must be dispatched at it's maximum output in at least one scenario \cite{snodgrass2021tractable}. Otherwise, the transmission lines connected to the generator points of interconnection (POI) for decommitted generators will be inadequately designed. 

To obtain a varied set of power injections from the generators, we create two sets of generation schedules for each hour using two methods, which we refer to as the \emph{economic dispatch} and \emph{uneconomic dispatch}. The economic and ``uneconomic'' dispatch scenarios result in all generators being dispatched at, or near, their maximum power output in at least one generator unit commitment scenario. 
\begin{itemize}
    \item \emph{Economic dispatch:} For each hour, we implement a simple unit commitment algorithm. This algorithm iteratively decommits the most expensive generator until a target spinning reserve of 10\% is met. Once the unit commitment is fixed, the generators are dispatched using an economic dispatch algorithm that minimizes the generator cost subject to the total demand equaling total generation.
    \item \emph{Uneconomic dispatch:} In the uneconomic dispatch, we follow a similar procedure, but instead decommit the cheapest generators to create an ``uneconomic dispatch''. The final injections are again computed by running an economic dispatch algorithm considering just the most expensive generators. %
\end{itemize}
Since we create two generation scenarios for each of our 245 load scenarios, we consider a total number of 490 power injection scenarios. %

\subsubsection{Initial Transmission Line and Transformer Parameters}
\label{sec:initial-TL-XF-Parameters}
As a final input to our method, initial transmission line and transformer impedances are assigned to each of the network branches. Instead of using a simple assignment such as assigning a uniform per-unit-length impedance to all transmission lines, we use transmission line data from the FERC Form 1 to make the initial assignment. For each transmission line at each voltage level in the CATS, we identify the line in the FERC Form 1 with the closest length. If the utility company listed in the CEC data matches the utility company listed in the FERC Form 1, we only examine the lines in the Form 1 data corresponding to that utility company. 

We used the Form 1 data for the matched CEC transmission lines, including conductor size (in kcmil), conductor type, and number of conductors per phase to determine ampacity limits and transmission line impedances for the corresponding transmission line in the CATS, 
following the methodology described in \cite{snodgrass2021tractable}.
As part of this process, we use transmission line manufacturer's data sheets \cite{ACSR2012} to determine ampacity limits and \cite{LaForest1982} to determine approximate geometric mean radius (GMR) and geometric mean diameter (GMD) values, which we then use to compute synthetic per-unit length transmission line impedances for the lines. %
We multiply the calculated ampacity limits by the rated voltage of the transmission line to calculate probable MVA thermal ratings for each transmission line in the Form 1.

By examining geographic regions such as states or approximate ISO or RTO service territory, we can determine MVA ranges for each voltage level and region. 
We validate the data using the MVA limit ranges for transmission lines at each voltage level in \cite{BirchfieldSchweitzer2017, snodgrass2021tractable}. We combine the calculated per-unit-length impedance parameters with the MVA limit ranges to produce a table of possible conductors configurations for each transmission line. We create the MVA values and ranges with the assumption that all conductors are aluminium conductor, steel-reinforced (ACSR) \cite{ACSR2012}. Later, we use this table to adapt the transmission line parameters as the MVA limits of lines are increased or decreased, as described below.

Since the Form 1 does not contain useful transformer data, we assign an initial limit of 2000 MVA to each transmission-level transformer. We obtain the average per unit impedance values from \cite{snodgrass2021tractable} using the transformer base MVA and the primary-secondary voltages. Corresponding X/R ratios are from \cite{BirchfieldSchweitzer2017}.
Once we calculate the initial line and transformer parameters, we solve a DC PF for all 490 scenarios to compute the resulting flows through all transmission lines and transformers. We then resize the transformers to have an MVA limit equal to the maximum value calculated from the DC PF and recompute the impedance parameters corresponding to these calculated flows. 

\subsection{Algorithm for Updating Transmission Line Parameters}
\label{sec:Updating-Parameters}

The initial generation schedules and line parameters, which we derive without accounting for the network constraints, often do not allow for feasible PF and OPF solutions. In the following section, we describe our algorithm for adjusting line parameters and generator dispatch to obtain feasible solutions.

\subsubsection*{\underline{Step 0: Initialize}} We define the 490 power injection scenarios and initial line parameters as discussed above.

\subsubsection*{\underline{Step 1: Solve line upgrade optimization problem}}
For each power injection scenario, we solve optimization problem \eqref{prob:dcopf}. 
\begin{subequations}
\label{prob:dcopf}
\begin{align}
    \label{eq:line-opt-obj}
    \min \quad & \lambda \sum_{k \in \mathcal{G}} \Delta P_{g,k}^s + (1-\lambda)\sum_{(i,j) \in \mathcal{L}} \delta_{ij}^s\\
    \label{eq:nodal balance}
    \mbox{s.t.} \quad &  P_{g,k} - \sum_{(i,j) \in \mathcal{L}} \beta_{ij}^kP_{f,ij} = P_{d,k} \quad \forall k \in \mathcal{B} \\ 
    \label{eq:line flows}
    & P_{f,ij} = -B_ij(\theta_i - \theta_j) \quad \forall (i,j) \in \mathcal{L} \\
    \label{eq:line limits}
    & -P_{f,ij}^{max} - \delta_{ij}^s \leq P_{f,ij} \leq P_{f,ij}^{max} + \delta_{ij}^s \quad \forall (i,j) \in \mathcal{L} \\
    \label{eq: redispatch gen}
    & P_{g,o}^s - \Delta P_{g,k}^s \leq P_{g,k}^s \leq P_{g,o}^s + \Delta P_{g,k}^s \quad \forall k \in \mathcal{G} \\
    & P_{g,k}^{min} \leq P_{g,k}^s \leq P_{g,k}^{max} \quad \forall k \in \mathcal{G} \\
    & \delta_{ij}^s \geq 0 \quad \forall (i,j) \in \mathcal{L} \\
    & \Delta P_{g,k}^s \geq 0 \quad \forall k \in \mathcal{G}
\end{align}
\end{subequations}
Sets $\mathcal{B}$ and $\mathcal{G}$ are the set of all the buses and generators in the grid respectively. The equality constraints \eqref{eq:nodal balance} represent the DC PF equations, while \eqref{eq:line limits} represent the relaxed transmission line limits and \eqref{eq: redispatch gen} represent the generation limits after redispatch. The primary outputs of this optimization problem are the line limit violations $\delta_{ij}^s$ for each line $(i,j) \in \mathcal{L}$ in each scenario $s \in\mathcal{S}$. Note that if the DC PF solution is feasible for the original power injection scenario $s$, the optimization problem would set $\delta_{ij}^s=0$. 
This optimization problem attempts to minimize the size of transmission line violations while also limiting generation redispatch away from the assigned power injection schedule. To achieve this, we formulate the objective function \eqref{eq:line-opt-obj} with two terms: (1) a penalty on $\Delta p_{g,s}$, which measures how much generator $g$ is redispatched in scenario $s$, and (2) a penalty on $\delta_{ij,s}$, which measures the violation of the PF limit on line $ij$ in scenario $s$. The factor $\lambda$ is a trade--off parameter that balances how much we penalize the generation redispatch and line limit violations in the solutions. A smaller value for $\lambda$ allows for more generation redispatch and leads to fewer line updates. A larger value for $\lambda$ penalizes generation redispatch more and thus forces more line upgrades. If we do not allow any redispatch at all, the procedure becomes similar to solving a PF for each load scenario. Based on testing with several values, we set $\lambda=0.5$ for our final grid. This results in a reasonable trade--off between upgrading transmission lines and limiting generation redispatch.

\subsubsection*{\underline{Step 2: Identify and upgrade overloaded lines}}  \label{sec:overload}
We compute the maximum violation across all scenarios to identify the overloaded lines,
\begin{equation}
    \delta_{ij} = \max_{s\in\mathcal{S}}~\delta_{ij,s}~.
\end{equation}
We randomly choose a subset of 509 overloaded lines to upgrade, or 5\% of the number of lines in the system\footnote{We upgrade only a subset of lines as some overload problems may be resolved in the next iteration once the other lines have been upgraded. The random choice of lines to upgrade reflects the fact that the power system has been evolving over a long period of time, and thus is not always built to be optimal for the present day loading conditions.}. If fewer than 509 lines are overloaded, we upgrade all overloaded lines. 

For each line $(i,j)$ that is chosen for an upgrade, we use the following procedure:
\begin{enumerate}
    \item[(a)] Using the table of possible conductor types and MVA ratings for this line, we upgrade the type of conductor to the one with the closest higher MVA rating. 
    \item[(b)] If the conductor type has already been upgraded to the highest MVA conductor type and could not be updated using the procedure in (a), we increase the number of circuits included in the line by one. Note that the maximum number of allowable circuits per line is 8. 
\end{enumerate}
If the line has already been upgraded to have 8 circuits, it is left in its overloaded state until the end of the algorithm. %

\subsubsection*{\underline{Step 3: Identify and upgrade underutilized lines}} 
We perform a similar set of changes to reduce the ratings of underutilized lines. 
Transmission lines with a utilization lower than a 30\%  are classified as underutilized. As done in the case of overloaded lines, a random subset of 509 underutilized lines are chosen. %
If the number of underutilized lines is smaller than 509, all such lines are downsized. 

For each of the chosen lines $(i,j)$:
\begin{enumerate}
    \item[(a)] Using the table of possible conductor types and MVA ratings, we downsize the line by choosing the conductor type that has the closest lower MVA rating
    \item[(b)] If the conductor type has already been modified to one with the lowest MVA rating, we reduce the number circuits in the line by one. Since a line must have at least one circuit, we do not decrease the number of circuits in a line once it equal one.
\end{enumerate}
If a line has already been downsized to have just the one circuit of the smallest allowable conductor size and it is still underutilized, we do not downsize it further.

\subsubsection*{\underline{Step 4: Check termination criterion}}
If the number of overloaded lines is below a threshold $\tau$, then the line resizing terminates. Otherwise, we return to solving the optimization problem in Step 1. 
At the beginning, the threshold $\tau$ is set to zero. However, if the algorithm fails to terminate after a certain number of iterations, the threshold is increased every iteration until the algorithm terminates. This prevents ``cycling'', where the algorithm upgrades a set of lines that causes another set of lines to become underutilized. Correcting these underutilized lines then causes the first set of lines to become overloaded, thus driving the algorithm into an infinite loop if the threshold $\tau$ is not increased.

\section{Assigning Reactive Power Support}
\label{sec:Reactive-Power}
The reactive power output of the generators alone is not sufficient to maintain the voltage at each bus within its limits. Thus, to ensure that the grid gives rise to an AC PF feasible solution where all voltage limits are satisfied, we add reactive power compensation elements to the network using the algorithm described below. Due to the high computational burden associated with solving AC OPF for a network of this size, this algorithm considers only a single power injection scenario corresponding to the maximum load scenario with economic generation dispatch. Before assigning reactive power compensation to the system, we temporarily double the thermal limits of all the lines to ensure that the reactive power flow and losses in the network, which were neglected when assigning line parameters, do not result in an infeasible AC OPF. The doubling of the thermal limits can be understood as changing the conductor type from ACSR to aluminum conductor, steel supported (ACSS), which roughly doubles the ampacity without a substantial change to the GMR of the wire \cite{ACSS2017}. This allows the MVA rating of a transmission line to increase by up to 100\% of the original value without modifying the corresponding impedance (R, X, and B) values.

\subsubsection*{\underline{Step 0: Initialize}} We add reactive power compensation in the form of synchronous condensers to all nodes in the network. The initial maximum capacity of the reactive power compensation devices is set to 200 MVAr. 

\subsubsection*{\underline{Step 1: Solve AC OPF}} For the problem with reactive power compensation installed, we solve a standard AC OPF problem which minimizes generation cost subject to AC PF, generation, transmission and voltage constraints \cite{Powermodels2018}.

\subsubsection*{\underline{Step 2: Remove redundant compensation}} We remove reactive power compensation from 20\% of the nodes that currently have reactive power compensation.

\subsubsection*{\underline{Step 3: Check termination criterion}} If fewer than 20\% of all nodes have reactive power compensation, then we terminate. Otherwise, we return to Step 1 and re-solve the AC OPF. This stopping criteria is based on the percentage of substations in the FERC Form 715 that have reactive power compensation \cite{FERC_Form715}. 

\subsubsection*{\underline{Step 4: Restore line limits}}
After assigning reactive power support, we restore the thermal limits of lines with a utilization of less than 50\% back to their original values (before they were doubled). As a result, 1.63\% of lines have limits that remain doubled (i.e., these lines are upgraded from ACSR to ACSS conductors).

\begin{figure}[t]
    \centering
    \includegraphics[width = \linewidth]{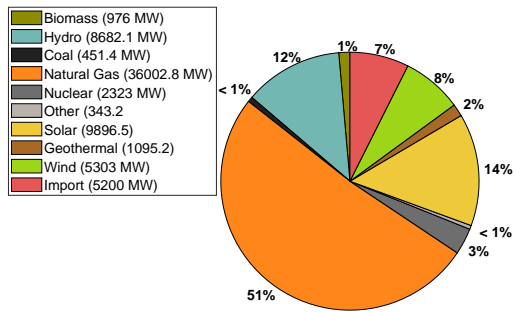}
    \caption{\small Percentage of generation from each fuel type.}
    \label{fig:fuel_types}
\end{figure}

\section{Results: Grid Metrics and Evaluation}
\label{sec:Metrics-Evaluation}
The  synthetic network has 8,870 buses, out of which 1,743 have reactive power support in the form of synchronous condensers\footnote{In future work we plan to convert these to discrete or switched capacitors.}. There are 10,162 transmission lines and 661 transformers in the grid. The system has 2,472 load buses with a peak load of 44,009 MW and 2,149 generators with a total capacity of 73,172 MW. Generation capacities by fuel type are shown in Fig. \ref{fig:fuel_types}. 

\begin{figure}[t]
    \centering
    \includegraphics[width = \linewidth]{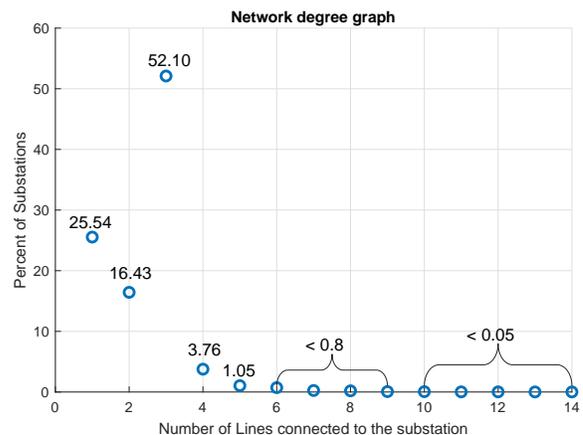}
    \caption{\small Node degree distribution for the network.}
    \label{fig: node_dist}
\end{figure}

An important metric in evaluating synthetic networks is the node degree distribution, which captures the frequency of the node degree (or the number of lines connected) at each substation. Fig. \ref{fig: node_dist} shows the node degree distribution for our synthetic grid. The node degree distribution of our network agrees closely with real networks, as there is a general downward trend with a peak between 2 and 3 lines per substation \cite{snodgrass2021tractable}.

Characteristics of the network branches, shown in Table \ref{tab:branch_stats}, follow trends similar to those of real grids presented in \cite{snodgrass2021tractable}. Any statistical comparison of the synthetic network and real grids will have shortcomings due to the fact that there are only three samples of real world networks in the United States.

\begin{table}[]
    \centering
    \caption{Characteristics of the network branches}
    \label{tab:branch_stats}
    \begin{tabular}{m{10mm} m{10mm} m{10mm} m{11mm} m{10mm} m{11mm}}
         \multirow{2}{3em}{\textbf{Voltage Levels}} & \multirow{2}{4em}{\textbf{Percent of lines}} & \multicolumn{2}{c}{\textbf{Length (Miles)}} & \multicolumn{2}{c}{\textbf{GVA-Miles}} \\
         \cline{3-6} 
         &  & \textbf{Form 1} & \textbf{CATS} & \textbf{Form 1} & \textbf{CATS}\\
         66 & 64.64 & 13124 & 12683.29 & 1070 & 1011.62\\
         115 & 21.74 & 10320 & 8076.95 & 2275 & 1725.07\\
         230 & 12.41 & 8296.81 & 10290 & 5736 & 5329.83\\
         500 & 1.21 & 4637.42  & 4428 & 10642 & 10083.78
    \end{tabular}
\end{table}

\begin{figure*}
    \centering
        \begin{subfigure}[t]{0.95\textwidth}
        \centering
        \includegraphics[width=1.0\textwidth]{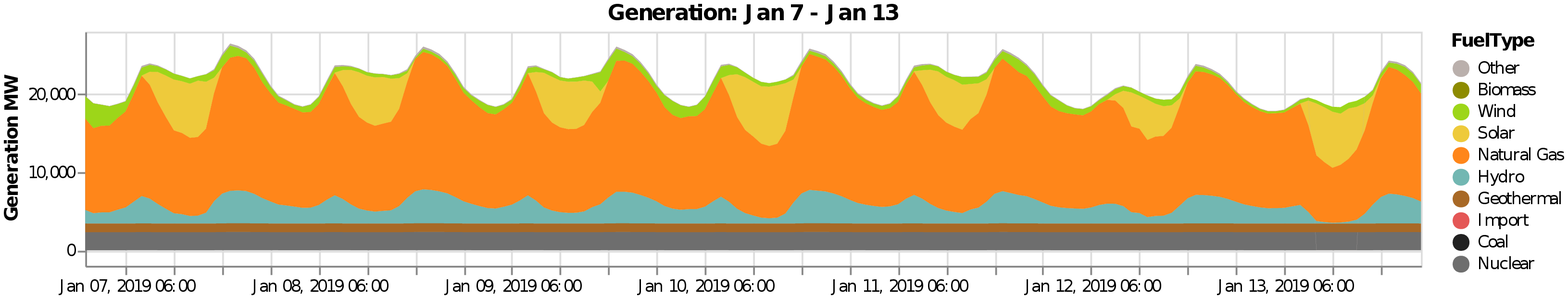} 
        \caption{} \label{fig:ac_winter}
    \end{subfigure}
    \begin{subfigure}[t]{0.95\textwidth}
        \centering
        \includegraphics[width=1.0\textwidth]{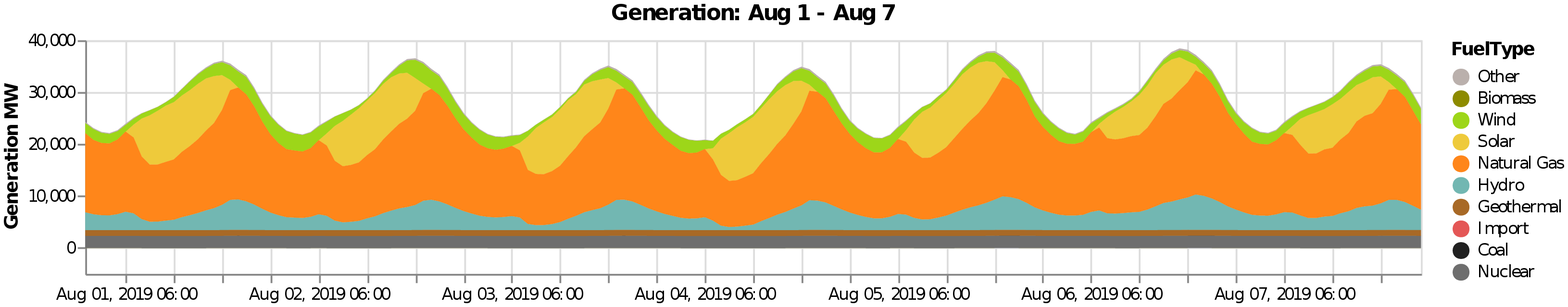} 
        \caption{}\label{fig:ac_summer}
    \end{subfigure}
    \caption{\small Generation dispatch from the AC solutions for a) Jan 7 through Jan 13. and b) Aug 1 through Aug. 7.
    }
    \label{fig:generation}
\end{figure*}

We evaluate the operation of the CATS grid across a year of hourly renewable generation and load scenarios to verify that the PF solutions are reasonable.
As described in Section \ref{sec:gen_load_data}, for each hour, we scale the capacity of each renewable generator according to its rating and the aggregate renewable production. For the load scenarios, we use a full year of the previously referenced hourly load data (see Section \ref{sec:data}). For each hour, we solve a DC and an AC OPF. These are independent problems and have no linking constraints between time periods, such as generator ramping constraints. 
Out of the 8,760 hourly time steps, each scenario is feasible and is solved to a locally optimal solution.

\begin{table}[t]
    \centering
    \caption{Congested transmission lines}
    \begin{tabular}{ccccc}
         \textbf{PowerFlow} &\textbf{Mean} & \textbf{Median} & \textbf{Maximum} & \textbf{Minimum} \\
         \hline
         DC-OPF & 2.68 &  1 & 25 & 0\\
         AC-OPF & 0.547 &  0 & 9 & 0
    \end{tabular}
    \label{tab:congestion}
\end{table}

\begin{table}[t]
    \centering
    \caption{Hourly cost }
    \begin{tabular}{ccccc}
         \textbf{PowerFlow} & \textbf{Mean} & \textbf{Median} & \textbf{Maximum} & \textbf{Minimum} \\
         \hline
         DC-OPF & \$629,000 &  \$623,000 & \$736,000 & \$499,000 \\
         AC-OPF & \$637,000 &  \$631,000 & \$747,000 & \$507,000
    \end{tabular}
    \label{tab:cost}
\end{table}

\begin{table}[t]
    \centering
    \caption{Hourly generation [MW]}
    \begin{tabular}{ccccc}
         \textbf{PowerFlow} & \textbf{Mean} & \textbf{Median} & \textbf{Maximum} & \textbf{Minimum} \\
         \hline
         DC-OPF & 22,574 &  21,807 & 39,709 & 15,255 \\
         AC-OPF & 23,202 &  22,412 & 41,157 & 15,829
    \end{tabular}
    \label{tab:generation}
\end{table}

We then evaluate aspects of the PF solutions, including feasibility, line loading, generation dispatch, and curtailment. %
Transmission line congestion is shown in Table \ref{tab:congestion}. Out of the 10,140 transmission lines in the network, there are on average 2.68 lines that are operating at their maximum capacity in the DC OPF solutions, while 0.547 on average are operating at their capacity on the AC OPF solutions. 
The highest congestion level in the DC OPF solutions occurs on August 15 at 7pm, when 25 lines are operating at their capacity. 
With AC OPF, a maximum of 9 lines are simultaneously at capacity. This occurs in 12 hours of the year, all between 5pm and 8pm in the months of June, July, August, and September. 
Overall, there is little congestion in this network in most hours.

The cost of operation is shown in Table \ref{tab:cost}. 
The operating cost with DC OPF is on average \$629,000 per hour of operation, but the maximum operating cost approaches \$736,000 and the minimum approaches \$499,000. 
The AC OPF solution typically costs 1.5\% more than the DC OPF solution. 
The hourly generation is shown in Table \ref{tab:generation}. 
The DC OPF problem does not contain losses, and the generation is equal to demand. 
The hourly generation ranges from 15,255 MWs to 39,709 MWs, with a average of 22,574 MWs of generation. 
The AC solution requires  approximately  3\% more generation to account for network losses. The hourly average cost of generating electricity is \$27 per MWh (\emph{note: this is not the marginal cost of generation}).

Generation profiles of the AC solutions are shown in Fig. \ref{fig:generation} for a winter profile (January 7 through 13) (Fig. \ref{fig:ac_winter}) and a summer profile (August 1 through 7)  (Fig. \ref{fig:ac_summer}). The generation output in the summer is much higher and has a larger ramp rate, as is expected for California. This is especially pronounced for natural gas (orange) after solar output (yellow) drops after sunset. We also note that the daily production period for solar energy is shorter in the winter than the summer, since there are fewer sunlight hours in the winter. 

Finally, we discuss the curtailment of wind and solar in the grid. 
The DC OPF solutions contain curtailment in 7 hours of the year, with a maximum of 5.44 MW of curtailed wind and solar. The AC OPF solutions have no curtailment in any hour of the year.
This does not reflect of the actual curtailment levels in the CAISO system.
The reason for this discrepancy might be that the hourly renewable energy data source is the dispatched power in the CAISO market, which is the amount of power \emph{after} curtailment has occurred. It may also be due to the fact that hourly OPF solutions do not include N-1 contingency, unit commitment, or generator ramping constraints between time periods, and introducing these constraints may cause renewable curtailment if some thermal generators cannot reduce their power output in order to meeting ramping requirements later in the day. Finally, there may be discrepancies between our grid model and the real grid.

\section{Conclusion}\label{sec:geo-conclusion}
The California Test System (CATS) is a geographically-accurate synthetic grid that can be used as a test case for policy-focused power systems research. The total transmission line lengths and capacities in CATS match closely with the real California grid. Additionally, the CATS grid has a feasible AC OPF solution for every hourly scenario in a year-long set of load data. 

To the our best knowledge, this is the first and only publicly available power grid model with accurate  geography. It is particularly valuable for use in applications that require geo-referenced grid data, such as those related to weather, climate change, topography, political boundaries, socio-economic considerations, and more. 
CATS is available in a GitHub repository \cite{githubRepo}.

We note that the final network is an approximation of California's transmission system. While the locations and paths of the components were not significantly modified, the connections and parameters are synthetic. This is important for maintaining security, but it also means that any results produced using this grid do not necessarily reflect the operation of California's actual grid.

In future development, we can improve the CATS grid by continuing to correct any remaining connectivity issues,
generating more realistic curtailment of renewable generators, and adjusting renewable generation limits based on local weather data. Our vision is that this test system will be a lasting tool for power systems research and policy development. We hope it will serve as a catalyst for designing future energy systems in harmony with complex environmental and social contexts.

\ifCLASSOPTIONcaptionsoff
  \newpage
\fi

\bibliographystyle{IEEEtran}
\bibliography{IEEEabrv, refs}

\end{document}